\documentstyle[11pt,aaspp4]{article}
\begin{document}

\title{Variable stars in the field of the globular cluster E3}

\author{B. J. Mochejska, J. Kaluzny}
\affil{Copernicus Astronomical Center, 00-716 Warszawa, Bartycka 18}
\affil{\tt e-mail: mochejsk@camk.edu.pl, jka@camk.edu.pl} 
\author{I. Thompson}
\affil{Carnegie Institution of Washington, 813 Santa Barbara Street,
Pasadena, CA 91101, USA}
\affil{\tt e-mail: ian@ociw.edu}
\begin{abstract}
We present the results of a search for variable stars in the faint
sparse globular cluster E3. We have found two variable stars: an SX
Phe variable (V1) and a W UMa eclipsing binary (V2).  We have applied
period-luminosity and period-color-luminosity relations to the
variables to obtain their distance moduli. V1 seems to be a blue
straggler belonging to E3, based on its distance modulus and location
on the CMD. V2 is probably located behind the cluster, in the Milky
Way halo.  We also present $V/B-V$ and $V/V-I$ color magnitude
diagrams of E3.
\end{abstract}

\section{Introduction}

We present the results of a search for variable stars in the faint
sparse globular cluster E3, located at $\alpha_{2000}=9^h20^m59^s$,
$\delta_{2000}=-77\arcdeg16\arcmin57\arcsec$. The cluster was
discovered on the ESO B Schmidt Survey of the Southern Sky by Lauberts
(1976). The first $BV$ photometry of the cluster was presented by van
den Bergh, Demers \& Kunkel (1980).  Numerous candidates for blue
stragglers were identified. The photoelectric photometry of Frogel \&
Twarog (1983) confirmed this finding. A subsequent study by Hesser et
al. (1984) using $UBV$ photoelectric and photographic observations
showed a sparsely populated subgiant branch in the color-magnitude
diagram.  The first CCD photometry for E3 in the $BV$ bands, obtained
with the CTIO 4m telescope, was published by McClure et
al. (1985). These observations suggested the presence of a second
sequence of stars $\sim0.75$ mag. above the cluster main sequence,
interpreted as evidence for a significant population of binary stars
in E3. The cluster was further studied by Gratton \& Ortolani (1987),
who provided new $BV$ CCD photometry from the 2.2m telescope at ESO.

\begin{figure}[t]
\plotfiddle{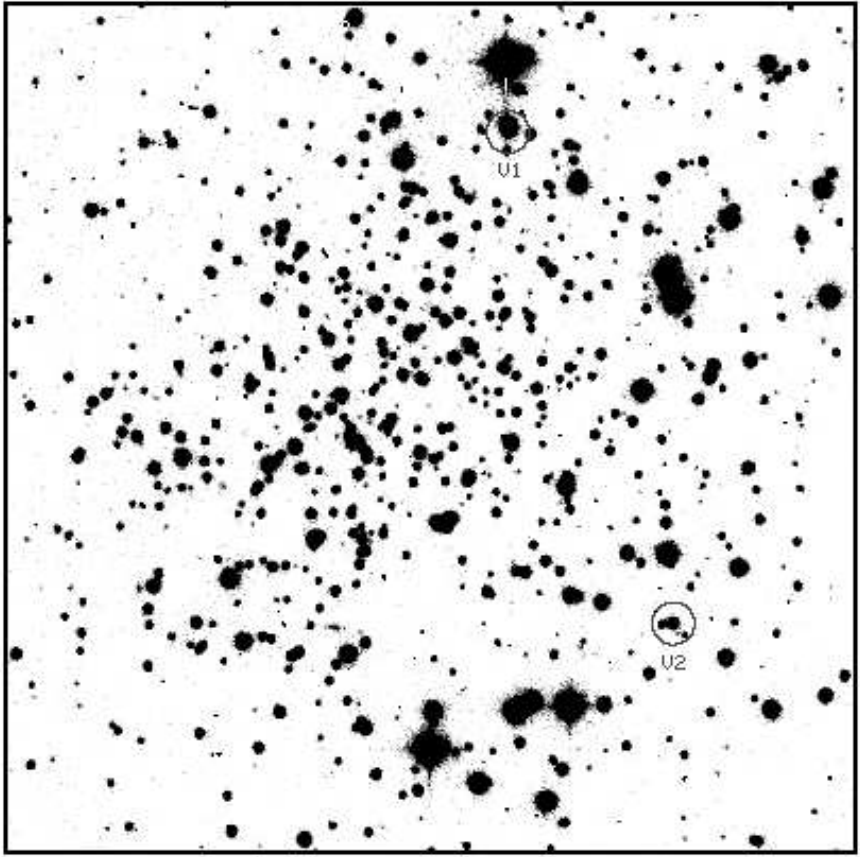}{10cm}{0}{80}{80}{-180}{-10}
\caption{Finding chart for the variables in E3. North is up and east
is to the left.}
\label{fig:map}
\end{figure}

\section{Observations and Data Reduction}

The data for this project was obtained with the Las Campanas
Observatory 1.0m Swope telescope during two separate runs: from April
11 to 21, 1996 and from May 16 to 27, 1996. For the first few nights
of the April run the telescope was equipped with the TEK1 1024x1024
CCD camera giving a pixel scale of $0.70\arcsec/pixel$. On the night
of April 14 the camera was switched to the TEK3 2048x2048 CCD with a
pixel scale of $0.61\arcsec/pixel$. During the May run the FORD
2048x2048 CCD camera with a pixel scale of $0.41\arcsec/pixel$ was
used.

The main observing target on both runs was the M4 globular cluster.
Several exposures of E3 were taken at the beginning of most nights.  A
total of 121 long ($400\div900\;sec$) exposures were taken in the $V$
filter (33, 42 and 46 with TEK1, TEK3 and FORD, respectively), six
short ($35\div120\;sec$) exposures in $V$ (2 with TEK1, 4 with FORD),
two $600\;sec$ exposures in $I$ (TEK1) and two $480\;sec$ exposures in $B$
(TEK1).

The preliminary processing of the CCD frames was done with the
standard routines in the IRAF-CCDPROC package.\footnote{IRAF is
distributed by the National Optical Astronomy Observatories, which are
operated by the Association of Universities for Research in Astronomy,
Inc., under cooperative agreement with the NSF} The images from the
TEK3 camera were clipped to a size of 1024x1024 pixels$^2$ to cover
roughly the same field as the TEK1 images. Due to the high degree of
psf variability on the images taken with the FORD camera, only the
central 800x800 pixel$^2$ sections were used.

\begin{figure}[!t]
\plotfiddle{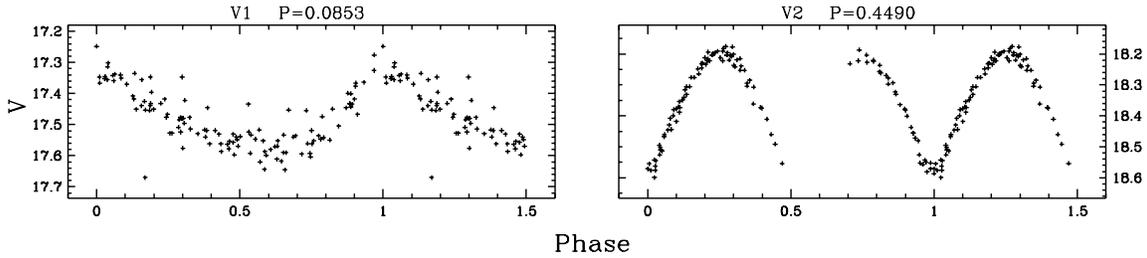}{2.3cm}{0}{80}{80}{-250}{-510}
\caption{Phased $V$ filter light curves of the two variables in E3.}
\label{fig:var}
\end{figure}

Photometry was extracted using the {\it Daophot/Allstar} package
(Stetson 1987). A PSF varying linearly with the position on the frame
was used. The PSF was modeled with a Moffat function. Stars were
identified using the FIND subroutine and aperture photometry was measured
with the PHOT subroutine. Approximately 40 bright isolated stars were
initially chosen by {\it Daophot} for the construction of the PSF.  Of
those the stars with profile errors greater than twice the average
were rejected and the PSF was recomputed. This procedure was repeated
until no such stars were left on the list. The PSF was then further
refined on frames with all but the PSF stars subtracted.
This procedure was applied twice. The PSF obtained in the above
method was then used by {\it Allstar} in profile photometry.

The image where the most stars were identified was chosen as the
template. The template star list was then transformed to the $(X,Y)$
coordinate system of each of the frames and used as input to {\it
Allstar} in the fixed-position mode. The output profile photometry was
transformed to the common instrumental system of the template image
and then combined into a database. The databases were created for the
long ($400-900\;sec$) exposures in the $V$ filter only.

\section{Variable stars}

We have followed the procedure for selecting variables given in
Kaluzny et al. (1998), where it is described in detail. From the 1541
stars in the $V$ database 11 variable star candidates were selected.
After the rejection of stars with noisy and/or chaotic light curves we
were left with two variables. Their periods were refined using the
analysis of variance method, as described by Schwarzenberg-Czerny
(1989). These two variables were confirmed with ISIS - the image
subtraction package (Alard \& Lupton 1998, Alard 1999). No other
variables were detected using this method.

In Figure \ref{fig:var} we present the phased $V$ light curves of the
two variables. Table \ref{tab:var} lists the parameters of these 
variables: name, period, $V$ magnitude ($\langle V\rangle$ for the 
pulsating variable, $V_{max}$ for the eclipsing binary), the $B-V$ and
$V-I$ colors. The variables are indicated by open circles on the finding
chart in Figure \ref{fig:map}.

V1 is a pulsating variable, most likely of the SX Phe type, judging
from its short period (0.0853 days) and the shape of its light
curve. V2 is an eclipsing binary with a period of 0.4490 days. Its
light curve shows an absence of the constant light phase, indicating
that it is a W UMa type variable.

We have used the period-luminosity calibration for SX Phe stars derived
by McNamara (1997) to estimate the distance modulus to V1:
\[M_V=-3.725 \log P - 1.933\]
Adopting a value of reddening E(B-V)=0.30 (Harris 1996) we obtain a
distance modulus $(m_V-M_V)_0=14.46$. This value is in agreement with
the distance moduli found in literature: 14.55 - van den Bergh et
al. (1980), 14.4 - Frogel \& Twarog (1983), 14.2 - Gratton \& Ortolani
(1987), indicating that V1 is located at the same distance as the
cluster.

The following period-color-luminosity calibrations for W UMa type
eclipsing binaries derived by Rucinski (2000) were applied to V2:
\[M_V^{BV} = -4.44 \log P + 3.03 (B-V)_0 + 0.12\]
\[M_V^{VI} = -4.43 \log P + 3.63 (V-I)_0 - 0.31\]

The fact that $B-V$ and $V-I$ colors were determined at random phase
should not influence the outcome substantially, as in the case of
contact binaries the color does not change significantly throughout
the cycle. Using a value of $E(V-I)=1.28 E(B-V)$ (Schlegel et
al. 1997) we obtained a distance modulus of 15.42 mag. from the first
calibration and 14.83 mag. from the second.  The variable appears to
be located behind the cluster, in the Milky Way halo.

\begin{table}
\caption[]{\sc Variables in E3\\}
\begin{tabular}{cccccl}
\hline\hline
Name& $P$ (days) & $V$ & $B-V$ & $V-I$ & Comments\\ 
\hline
V1  & 0.0853 & 17.48 & 0.60 & 0.86 & SX Phe\\
V2  & 0.4490 & 18.17 & 0.66 & 0.96 & W UMa \\
\hline
\end{tabular}
\label{tab:var}
\end{table}

\section{Color-magnitude diagrams}
To construct the color-magnitude diagrams we combined pairs of long
exposures in the $BVI$ filters. 
The transformation from instrumental magnitudes to the standard system 
was derived from the observations of the Landolt fields (Landolt 1992).
The following relations were adopted:
\begin{eqnarray*}
v = V - 0.0189 \times (B-V) + const\\
b - v = 0.9359 \times (B-V) + const\\
v = V - 0.0182 \times (V-I) + const\\
v - i = 0.9843 \times (V-I) + const
\end{eqnarray*}

We have compared our $BV$ photometry with that of McClure et al. (1985).
The average differences in $V$ magnitude and $B-V$ color were computed
for 6 selected stars in the range $15.5\leq V\leq 19.25$ and were found to
be $\Delta V=0.02\pm0.014$ and $\Delta (B-V)=0.04\pm0.080$.

\begin{figure}[tb]
\vspace{7cm}
\includegraphics{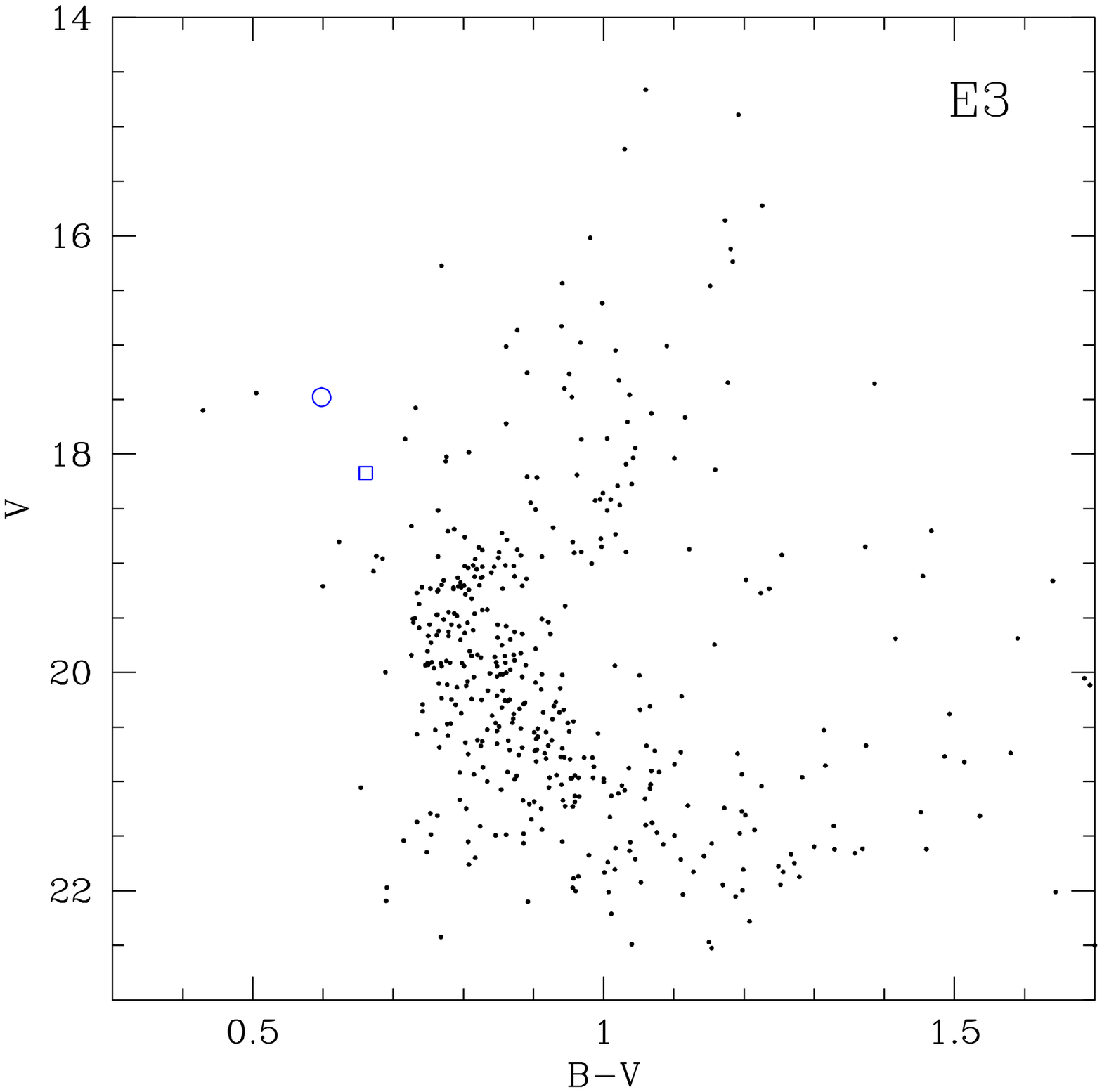}
\includegraphics{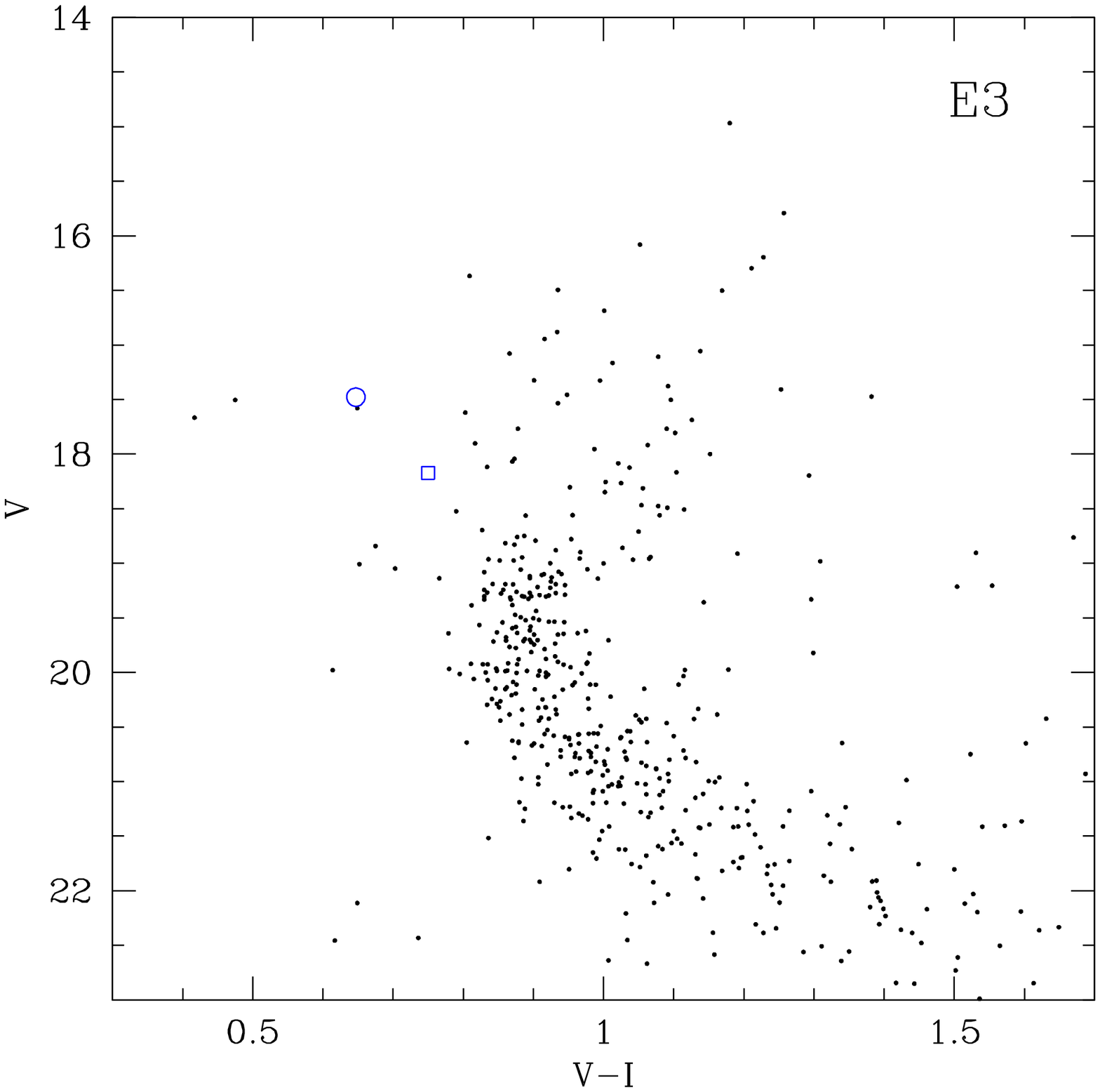}
\caption{The $V/B-V$ and $V/V-I$ color-magnitude diagrams of the inner
$2\arcmin$ of the E3 globular cluster. Variable V1 is denoted by an open
circle and V2 by an open square.}
\label{fig:cmd}
\end{figure}

The colors and magnitudes for variable stars were determined following
a different procedure. For variable V1 its average magnitude $\langle
V\rangle$ was used to place it on the CMDs. V2 was plotted with its
magnitude outside of the eclipses $V_{max}$. To derive the colors we
used the single $B$ and $I$ exposures and interpolated the $V$
magnitudes from the nearest exposures in $V$ to those epochs. The
final values of $B-V$ and $V-I$ were taken as the average of two color
determinations, with the average scatter of 0.04 mag.

The resultant $V/B-V$ and $V/V-I$ color-magnitude diagrams are shown
in Figure \ref{fig:cmd}, with the variable V1 denoted by an open
circle and V2 by an open square. Only stars within $2\arcmin$ of the
cluster center are plotted. Both variables are located among candidate 
blue stragglers, although V2 appears to be located behind the cluster,
based on the distance modulus determination in the previous section.

The cluster main sequence is apparent in both diagrams.  It exhibits
considerable scatter and there is some indication of a second sequence
running above it, although not as clear as in Figure 3 of McClure et
al. (1985). This would indicate that E3 could possess a significant
population of binary stars. This is in agreement with the idea
proposed by van den Bergh et al. (1980) that severe tidal stripping
had depleted the cluster in single stars, leading to an increased
binary frequency. 

A gap in the main sequence near the turnoff, at $V\sim19.5$ is visible
in both CMDs.  This has been previously noted by McClure et al. (1985)
and shown to be more of a visual effect, as no significant
discontinuities are present in the cumulative luminosity function for
stars on the main sequence (Figure 6 therein). This result is
confirmed by our analysis.

The subgiant and lower giant branches are also discernible in the
diagrams, although they show substantial scatter. This is regarded as
a real feature of the cluster, as commented in literature
(i.e. Hesser et al. 1984). A number of stars blueward of the turnoff
are present, possibly blue stragglers belonging to the cluster, as
first noted by van den Bergh et al. (1980).

\section{Conclusions}

Our variability search in E3 resulted in the discovery of two variable
stars: an SX Phe variable (V1) and a W UMa eclipsing binary (V2).  We
have applied period-luminosity and period-color-luminosity relations
to the variables to obtain their distance moduli. V1 seems to be a
blue straggler belonging to E3, based on its distance modulus and
location on the CMD. V2 is probably located behind the cluster.

\acknowledgments{We would like to thank Krzysztof Z. Stanek for his
error scaling, database manipulation and period finding programs and
Grzegorz Pojma{\'n}ski for $lc$ - the light curve analysis utility,
incorporating the analysis of variance algorithm. BJM and JK were
supported by the polish KBN grant 2P03D003.17 and by NSF grant
AST-9819787.}


\begin{references}
\reference{} Alard, C. 1999, preprint (astro-ph/9903111)
\reference{} Alard, C., Lupton, R.~H. 1998, ApJ, 503, 325
\reference{} Frogel, J.A., Twarog, B.A. 1983, ApJ, 274, 270
\reference{} Gratton, R.G., Ortolani, S. 1987, A\&AS, 67, 373
\reference{} Harris, W.~E. 1996, AJ, 112, 1487
\reference{} Hesser,J.~E., McClure, R.~D., Hawarden, T.~G., Cannon, R.~D., 
             von Rudloff, R., Kruger, B., Egles, D. 1984, PASP, 96, 406
\reference{} Kaluzny, J., Stanek, K.~Z., Krockenberger, M., Sasselov, D.~D.,
             Tonry, J.~L., \& Mateo, M. 1998, AJ, 115, 1016
\reference{} Landolt, A. 1992, AJ, 104, 340
\reference{} Lauberts, A. 1976, A\&A, 52, 309
\reference{} McClure, R.~D., Hesser, J.~E., Stetson, P.~B., Stryker, L.~L.
             1985, PASP, 97, 665
\reference{} McNamara, D.~H. 1997, PASP, 109, 1221
\reference{} Rucinski, S.~M. 2000, preprint (astro-ph/0001152)
\reference{} Schlegel, D. J., Finkbeiner, D. P., Davis, M. 1997, AAS,
             191, 8704
\reference{} Schwarzenberg-Czerny, A. 1989, MNRAS 253, 198
\reference{} Stetson, P.~B. 1987, PASP, 99, 191
\reference{} van den Bergh, S., Demers, S., Kunkel, W.~E. 1980, ApJ, 239,
             112
\end{references}
\end{document}